\documentclass{PoS}
\usepackage{amsfonts} 
\usepackage{amssymb} 
\usepackage{amsmath} 
\usepackage{graphicx} 
\usepackage{array} 
\usepackage{bm} 
\graphicspath{{./figs/}}

\title{Non-Goldstone pion masses with NLO in Staggered Chiral Perturbation Theory}

\ShortTitle{Non-Goldstone pion masses}

\author{Jon A.~Bailey, \speaker{Hyung-Jin Kim}, and Weonjong Lee \\
  Lattice Gauge Theory Research Center, FPRD, and CTP \\
  Department of Physics and Astronomy, 
  Seoul National University, Seoul, 151-747, South Korea \\
  E-mail: \email{jabsnu@gmail.com}, \email{windy510@gmail.com}, 
  \email{wlee@snu.ac.kr}}

\abstract{
  We present results of the masses of taste non-Goldstone $(F \ne
  \xi_5)$ pions and kaons calculated up to the next-to-leading order
  in the SU(3) staggered chiral perturbation theory (SChPT).
  The results can be used to fit data and to understand taste symmetry
  breaking effect quantitatively.
  The final expressions for the non-Goldstone masses contain 20 low
  energy constants unique to the non-Goldstone sector.
  We have calculated the several cases such as the full QCD, partially
  quenched QCD, and quenched QCD in the $N_f=1+1+1$ flavor and
  $N_f=2+1$ flavor cases in the $SU(3)$ and $SU(2)$ SChPT.
  In this paper, we present only the SU(3) part. 
}

\FullConference{ The XXIX International Symposium on Lattice Field Theory 
  - Lattice 2011\\
  July 10-16, 2011\\
  Squaw Valley, Lake Tahoe, California}

\begin{document}
%
\newcommand{\TB}{\textrm{B}}
\newcommand{\TI}{\textrm{I}}
\newcommand{\TP}{\textrm{P}}
\newcommand{\TV}{\textrm{V}}
\newcommand{\TA}{\textrm{A}}
\newcommand{\TT}{\textrm{T}}
\newcommand{\CC}{{\cal C}}
\newcommand{\CD}{{\cal D}}
\newcommand{\CF}{{\cal F}}
\newcommand{\CL}{{\cal L}}
\newcommand{\CO}{{\cal O}}
\newcommand{\CS}{{\cal T}}
\newcommand{\CT}{{\cal T}}
\newcommand{\CV}{{\cal V}}
\newcommand{\CU}{{\cal U}}
\newcommand{\CM}{{\cal M}}
\newcommand{\dmu}{\partial_\mu}
\newcommand{\tr}{\textrm{tr}}
\newcommand{\Tr}{\textrm{Tr}}
\section{Introduction}
Lattice calculations using improved staggered fermions have been making
a steady progress in particle physics of hadrons \cite{milc-rmp-1}.
Improved staggered fermions such as HYP and HISQ have a number of
merits such as an exact chiral symmetry in the Goldstone taste channel
($\xi_5$), very small taste symmetry breaking in the non-Goldstone
taste channel, the scaling violation under control and so on
\cite{wlee-2006-1}.
In order to analyze the data calculated using staggered fermions,
staggered chiral perturbation theory (SChPT) has been developed
\cite{Lee:1999zxa,Aubin:2003mg,Sharpe:2004is}.
In Ref.~\cite{Aubin:2003mg}, Aubin and Bernard has used SChPT to
calculate the masses of pions and kaons (flavor-charged states) in the
Goldstone pion sector.
Here we extend their work to the non-Goldstone pion sector.  
The results will facilitate more precise determination of the light
quark masses, Gasser-Leutwyler couplings, and related quantities.
\section{Staggered effective chiral Lagrangian}
We can construct the SChPT lagrangian out of the PGB (pseudo-Goldstone
boson) fields $\phi$, quark mass matrix $M$, derivatives, and taste
matrices $\xi_\mu$ according to the symmetries of staggered fermions.
We can express the staggered fermion action in terms of the Symanzik
effective action as in \cite{Lee:1999zxa,Aubin:2003mg,Sharpe:2004is}.
In this effective action, the taste symmetry of individual terms
are quite different from one another.
For example, the lowest order terms respects SO(4) taste symmetry,
while the higher order terms break this symmetry down to
$\text{SW}_4$.

The exponential parameterization is a convenient way to include the PGBs.
\begin{equation}
\Sigma \equiv e^{i\phi/f}, \qquad SU(12)_L\times SU(12)_R:
\  \Sigma\rightarrow L\Sigma R^\dagger 
\end{equation}
where $L,\ R\in SU(12)_{L,R}$, respectively and 
\begin{eqnarray}
  \phi=\sum_a \phi^a \otimes T^a \,,\quad T^a&\in&\{\xi_5,\ i\xi_{\mu5},
  \ i\xi_{\mu\nu}(\mu<\nu),\ \xi_\mu, I\}\,,\quad
  \phi^a={
\begin{pmatrix}
U_a & \pi^+_a & K^+_a \\
\pi^-_a & D_a & K^0_a \\
K^-_a & \bar K^0_a & S_a
\end{pmatrix}}
\end{eqnarray}
The index $a$ runs over the 16 PGB tastes in the $\mathbf{15}$ and 
$\mathbf{1}$ of $SU(4)_T$, the $\phi^a$ are Hermitian $3\times 3$ matrices.
And the $T^a$ are Hermitian $4\times 4$ generators of $SU(4)_T$,
chosen as members of the Clifford algebra generated by the matrices
$\xi_\mu$.
With this choice for the $T^a$, the $SO(4)_T$ quantum numbers of the
PGBs are explicit.
At leading order in the expansion of the Lagrangian, there are three
classes of interactions: operators of ${\cal O}(p^2) \approx {\cal O}(m)
\approx {\cal O}(a^2)$.
We have
\begin{equation}
\mathcal{L}_\mathrm{LO} =\frac{f^2}{8} \Tr(\partial_{\mu}\Sigma \partial_{\mu}\Sigma^{\dagger}) - 
\frac{1}{4}\mu f^2 \Tr(M\Sigma+M\Sigma^{\dagger}) + \frac{2m_0^2}{3}(U_I + D_I + S_I)^2 
+ a^2 (\mathcal{U+U^\prime})
\label{F3LSLag}
\end{equation}
We follow the notations from Refs.~\cite{Aubin:2003mg,Sharpe:2004is}
in Eq.~\eqref{F3LSLag}.
The derivation of the potentials is described in
Refs.~\cite{Lee:1999zxa, Aubin:2003mg, Sharpe:2004is}.

At next-to-leading order(NLO), the Lagrangian operators fall into six
classes: $(n_{p^2},n_m,n_{a^2})=(2,0,0)$, $(0,2,0)$, $(1,1,0)$,
$(1,0,1)$, $(0,1,1)$, and $(0,0,2)$, where the $(n_{p^2},n_m,n_{a^2})$
vector represents terms of order ${\cal O}(p^{2n_{p^2}} m^{n_m} a^{2
  n_{a^2}})$.
The first three contain terms have an lattice analogy to the
Gasser-Leutwyler Lagrangian~\cite{Gasser:1984gg}.
The last three contain the terms are given in
Ref.~\cite{Sharpe:2004is}.

\section{\label{sec:se}Self-energies of $\pi^+$ mesons}
Here, we want to obtain the self-energy corrections up to NLO in SChPT.
In terms of the self-energy $\Sigma(p^2)$ of the state $\phi_{xy}^a\
(x\neq y,\ x,y\in\{u,d,s\})$,
\[
M_\phi^2= m^2_\phi + \Sigma(-M^2_\phi)=m^2_\phi + \Sigma(-m^2_\phi) - 
\Sigma(-M^2_\phi)\Sigma^\prime(-m^2_\phi) + \dots=m^2_\phi + \Sigma(-m^2_\phi) + \mathrm{NNLO} 
\]
where $m_\phi$ is the tree-level (LO) mass, and $M_\phi$ is the
(exact) mass to all orders.
Noting that expanding $\Sigma(p^2)$ in a Taylor series around
$p^2=-m^2_\phi$ gives, the NLO correction to the mass is the leading
contribution to $\Sigma(-m^2_{\phi})$.

\begin{figure}
\begin{center}
\includegraphics[width=10cm]{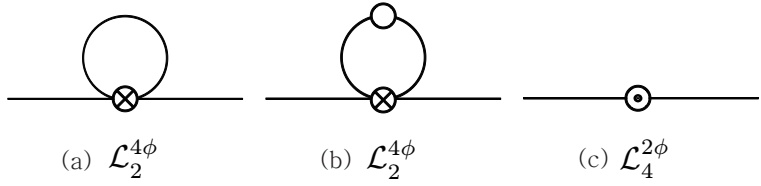}
\caption{ (a) The propagator represents the connected part.  (b)
  Disconnected tadpoles enter in the flavor-neutral, taste-singlet,
  taste-vector, and taste-axial channels.  (c) tree-level graphs
  contribute terms analytic in the quark masses and lattice spacing.
  The vertices are from the Gasser-Leutwyler and Sharpe-Van de Water
  Lagrangians }
\label{fig:scon}
\end{center}
\end{figure}

Expanding the LO Lagrangian through $\CO(\phi^4)$ and the NLO
Lagrangian through $\CO(\phi^2)$, we can obtain
\begin{equation}
\Sigma(p^2) = \frac{1}{(4\pi f)^2}\left[ \sigma^{con}(p^2)
 + \sigma^{disc}(p^2) \right] + \sigma^{anal}(p^2) + \dots\label{SelfEDiv}
\end{equation}
Where $\sigma^{con}$ is the sum of connected tadpole diagrams
(Fig.~\ref{fig:scon}(a)), $\sigma^{disc}$ is the sum of disconnected
tadpoles (Fig.~\ref{fig:scon}(b)), and $\sigma^{anal}$ is the sum of
analytic diagrams at the tree level (Fig.~\ref{fig:scon}(c)).

The 4-point vertices in the tadpole graphs are from the $\CO(\phi^4)$
terms in the LO Lagrangian of Eq.~(\ref{F3LSLag}).
The 2-point vertices in the tree-level diagrams are from the
$\CO(\phi^2)$ terms in the NLO Lagrangian of Gasser and Leutwyler, and
the NLO Lagrangian of Sharpe and Van de Water~\cite{Sharpe:2004is}.
And the disconnected propagators (in the graphs of
Fig.~\ref{fig:scon}(b)) are from quark-level disconnected
contributions to the tree-level, flavor-neutral propagators in the
taste singlet, axial, and vector channels~\cite{Aubin:2003mg}.

The one-loop graphs break taste $SU(4)_T$ to the remnant taste
$SO(4)_T$ of Ref.~\cite{Lee:1999zxa}.
The tree-level graphs from the Gasser-Leutwyler lagrangian respect
$SU(4)_T$, and the tree-level graphs from the Sharpe-Van de Water
lagrangian break $SU(4)_T$ in two stages:
Terms of $\CO(a^2m_q)$ and $\CO(a^4)$ break $SU(4)_T$ to $SO(4)_T$,
while terms of $\CO(a^2p^2)$ break the symmetry to $SW_\mathrm{3}
\subset SW_\mathrm{4} \subset SO(4)_T$~\cite{Sharpe:2004is}.
\subsection{\label{sec:pav}Propagators and vertex classes}
Here, we begin with the $4+4+4$ flavor-taste SChPT, which will
be reduced to the $1+1+1$ flavor theory later. 
Expanding the LO Lagrangian Eq.~(\ref{F3LSLag}) through $\CO(\phi^2)$ 
yields the propagators~\cite{Aubin:2003mg}.  
They are
\begin{equation}
\langle\phi^a_{ij}\phi^b_{k\ell}\rangle=
\delta^{ab}\left(\delta_{i\ell}\delta_{jk}
\frac{1}{q^2+m_{ija}^2}+\delta_{ij}\delta_{k\ell}D^a_{i\ell}\right)
\label{prop}
\end{equation}
where $i,j,k,\ell$ are flavor $SU(3)_F$ indices, $a,b$ are taste
indices in the adjoint irrep, and $m_{ija}$ is the pion mass with
flavors $i$ and $j$, and with taste $\xi_a$.
\begin{equation}
D^a_{i\ell}=-\frac{\delta_a}{(q^2+m_{iia}^2)(q^2+m_{\ell\ell a}^2)}
\frac{(q^2+m_{U_a}^2)(q^2+m_{D_a}^2)(q^2+m_{S_a}^2)}
{(q^2+m_{\pi^0_a}^2)(q^2+m_{\eta_a}^2)(q^2+m_{\eta^\prime_a}^2)}
\label{D_piece}
\end{equation}
where
\begin{equation}
\delta_I = 4m_0^2/3\,,\quad \delta_\mu = a^2\delta_V^\prime\,,\quad
\delta_{\mu\nu} = 0\,,\quad\delta_{\mu5} = a^2\delta_A^\prime\,,
\quad\delta_5 = 0
\label{deltas}
\end{equation}
\begin{equation}
m_{iia}^2 = 2\mu m_i + a^2\Delta_a \,,
\quad m_{ija}^2 = \mu(m_i+m_j)+a^2\Delta_a
\label{eq:tm}\,,
\end{equation}
where $m_i$ is the quark mass of flavor $i$.
$m_{\pi^0_a}^2$, $m_{\eta_a}^2$, and $m_{\eta^\prime_a}^2$ are the
eigenvalues of the matrix
\begin{equation}\label{444_mass_matrix}
	\begin{pmatrix}
	m^2_{U_a} +\delta_a & \delta_a & \delta_a \\
	\delta_a & m^2_{D_a} +\delta_a & \delta_a \\
	\delta_a & \delta_a & m^2_{S_a} +\delta_a
	\end{pmatrix},
\end{equation}
and the hairpin couplings $\delta_{V,A}^\prime$ (Eqs.~(\ref{deltas}))
and taste splittings $\Delta_a$ of the tree-level masses
(Eqs.~(\ref{eq:tm})) are defined in terms of the couplings of the LO
Lagrangian~\cite{Aubin:2003mg}.

Expanding the LO Lagrangian of Eqs.~(\ref{F3LSLag}) and keeping terms
of $\CO(\phi^4)$ gives 11 classes of vertices.
The kinetic term produces two classes, and the mass terms generate
one.
The potential $a^2\mathcal{U}$ gives  four classes, 
and the potential $a^2\mathcal{U}^\prime$ provides four classes.
As an example, the kinetic term gives the following contribution:
\begin{equation}
\frac{f^2}{8} \Tr(\dmu\Sigma \dmu\Sigma^{\dagger})=\phantom{-}\frac{1}{48f^2}
\,\tau_{abcd}\,(\dmu\phi^a_{ij}\,\phi^b_{jk}\,\partial_\mu\phi^c_{kl}\,\phi^d_{li}
-\dmu\phi^a_{ij}\,\dmu\phi^b_{jk}\,\phi^c_{kl}\phi^d_{li})+ \dots
\label{KE4pt}
\end{equation}
where the indices $a,b,c,d$ run over the 16 tastes in $SU(4)_T$, and
$\tau_{abc\cdots}\equiv\Tr(T^aT^bT^c\cdots)$ is the trace of products
of (Hermitian) taste matrices.

\subsection{\label{sec:vres}Results classified by vertex type}
Here, we want to present the results of Feynman diagrams after
contraction of loops and to classify them according to the vertex
type.
We consider external fields $\phi_{xy}^t$ and $\phi_{yx}^t$, where $t$
is the valence taste index, $x\neq y$ are the valence flavor indices,
and we use the renormalization scheme of Ref.~\cite{Aubin:2003mg}.
For the tadpole graphs with kinetic energy vertices
(Eq.~(\ref{KE4pt})), we find
\begin{equation}
I_\textrm{kin} = 
\frac{1}{12f^2}\sum_a\Biggl[p^2\Biggl(\sum_i (K_{xi,ix}^a + K_{yi,iy}^a) 
- 2\theta^{at}K_{xx,yy}^a\Biggr)+\sum_i(L_{xi,ix}^a + L_{yi,iy}^a) 
- 2\theta^{at}L_{xx,yy}^a \Biggr]
\label{KE444pre}
\end{equation}
where $i=u,d,s$ runs over the sea flavors in the loops, $a$ is the
taste of mesons in the loops,
$\theta^{ab}\equiv\frac{1}{4}\tau_{abab}=\pm1$ if $[T^a,T^b]_{\mp}=0$,
and
\begin{equation}
K_{ij,kl}^a\equiv\int \frac{d^4q}{(2\pi)^4}\langle\phi_{ij}^a\phi_{kl}^a\rangle\quad
\mbox{,}\quad L_{ij,kl}^a\equiv\int \frac{d^4q}{(2\pi)^4}q^2\langle\phi_{ij}^a\phi_{kl}^a\rangle.
\end{equation}
Substituting for the propagators and performing the integrals for the
connected contributions gives,
\begin{equation}
I_\textrm{kin} = 
\frac{1}{12f^2}\sum_a\Biggl[\frac{1}{(4\pi)^2}\sum_{Q}(p^2-m_{Q_a}^2)l(Q_a)+
\int\frac{d^4q}{(2\pi)^4}(p^2+q^2)(D_{xx}^a + D_{yy}^a - 2\theta^{at}D_{xy}^a)\Biggr]
\label{KE444}
\end{equation}
where $Q$ runs over the six valence-sea flavor combinations, 
and $l(X)\equiv m_X^2\ln m_X^2/\Lambda^2$.
In the same way, we can expand the diagrams with the mass vertices,
$\mathcal{U}$, and $\mathcal{U'}$.

The tree-level graphs with vertices from the Sharpe-Van de Water Lagrangian 
may be parameterized by introducing low-energy couplings corresponding 
to the irreps of $SO(4)_T$ and $SW_\mathrm{3}$:
\begin{equation}
-\frac{16}{f^2}a^2(A^tm_{xy5}^2 + B^t\,4(m_{U_5}^2+m_{D_5}^2+m_{S_5}^2)+ C^tp^2 + D^ta^2)\label{eq:SV}
\end{equation}
where the coefficients $A^t$, $B^t$, and $D^t$ are degenerate within
the $SO(4)_T$ irreps, and the coefficients $C^t$ are degenerate within
the $SW_\mathrm{3}$ irreps.

\subsection{\label{sec:res444}Results in 4+4+4 theory}
The results in Eq.~(\ref{KE444}) and those of the other cases (mass,
$\mathcal{U}$, $\mathcal{U'}$) are the one-loop contributions to the
expansion of the (negative of the) self-energies of the flavor-charged
PGBs of taste $t\in\{I,\xi_\mu,\xi_{\mu\nu}(\mu<\nu),\xi_{\mu5},
\xi_5\}$ in the 4+4+4 theory.
Collecting the connected contributions and factoring $-1/(4\pi f)^2$
and setting $p^2=-m_{xyt}^2=-m_{xy}^2-a^2\Delta_t$, we have
\begin{equation}
\sigma^{con}=-a^2\sum_B\Biggl(\delta_{BF}^{con}\,l(xy_B)+ 
\frac{\Delta_{BF}^{con}}{12}\sum_Q l(Q_B)\Biggr)
\label{fin_con444}
\end{equation}
\begin{equation}
\delta_{BF}^{con}\equiv\frac{1}{32}\sum_{a\in B}\sum_{b\in V,A}
\delta_b^{\prime+}\tau_{abt}\tau_{abt}(1+\theta^{ab}) \mbox{ ,}\quad
\delta_b^{\prime+}\equiv\frac{16}{f^2}
\begin{cases}
C_{2V}+C_{5V} & \text{if $b\in\{\mu\}$} \\
C_{2A}+C_{5A} & \text{if $b\in\{\mu5\}$}
\end{cases}
\end{equation}
\begin{equation}
\Delta_{BF}^{con}\equiv\sum_{a\in B}\biggl(\Delta_{at}-(\Delta_t+\Delta_a)\biggr)
\label{DelBFcon_def}
\end{equation}
where the indices $B$ and $F$ represent the taste of the loop meson
and the taste of the external valence taste, respectively.
The indices $B$ and $F$ are classified into the $SO(4)_T$ irreps;
$B,F\in\{I,V,T,A,P\}$; and $t\in F$.
\begin{table}[h]
\begin{tabular}{c || c | c || c || c | c | c }
\hline
B & F=V & F=A & B & F=T & F=P & F=I \\
\hline
\small{T} & \footnotesize{$3(C_{2A}+C_{5A}-C_{2V}-C_{5V})$} & \footnotesize{$3(C_{2V}+C_{5V}-C_{2A}-C_{5A})$} 
& \small{V} & \small{$2(C_{2A}+C_{5A})$} & 0 & \small{$4(C_{2V}+C_{5V})$}\\
\small{I} & $C_{2V}+C_{5V}$ & $C_{2A}+C_{5A}$& \small{A} & \small{$2(C_{2V}+C_{5V})$} & 0 & \small{$4(C_{2A}+C_{5A})$}\\
\hline
\end{tabular}
\caption{$\delta_{BF}^{con}$ coefficients for each pion and loop taste, multiply the coefficient $16/f^2$ }
\label{ConDiag2}
\end{table}
\begin{table}[h]
\begin{tabular}{c || c | c |  c | c | c}
\hline
B  & F=V & F=A & F=T & F=P & F=I\\
\hline
V & $4C_1+C_3+9C_4+6C_6$ & $4C_1+3C_3+3C_4+6C_6$ & $2C_3+6C_4+8C_6$ & $0$ & $4C_3+12C_4$\\
A & $4C_1+3C_3+3C_4+6C_6$ & $4C_1+9C_3+C_4+6C_6$ & $6C_3+2C_4+8C_6$ & $0$ & $12C_3+4C_4$\\
T & $3C_3+9C_4+12C_6$ & $9C_3+3C_4+12C_6$ & $6C_3+6C_4+16C_6$ & $0$ & $12C_3+12C_4$\\
P & $0$ & $0$ & $0$ & $0$ & $0$\\
I & $C_3+3C_4$ & $3C_3+C_4$ & $2C_3+2C_4$ & $0$ & $4C_3+4C_4$\\
\hline
\end{tabular}
\caption{The connected diagram, $\Delta_{BF}^{con}$ result. multiply the coefficient $6\times 16/f^2$ }
\label{ConDiag}
\end{table}

The coefficients $\delta_{BF}^{con}$ and $\Delta_{BF}^{con}$ are
linear combinations of the couplings in the potentials
$\mathcal{U^\prime}$ and $\mathcal{U}$, respectively.
Explicit results for $\delta_{BF}^{con}$ and $\Delta_{BF}^{con}$ are
given in Tables~\ref{ConDiag2} and \ref{ConDiag}.
We note that $\delta_{PF}^{con}=\Delta_{PF}^{con}=0$ which follows the
connected contributions in the Goldstone case.
\\

Collecting the disconnected pieces at the one-loop level in
Eq.~(\ref{KE444}) and the other cases(mass, $\mathcal{U}$,
$\mathcal{U'}$) gives
\begin{align}
\sigma^{disc}=&-\frac{(4\pi)^2}{12}\int\frac{d^4q}{(2\pi)^4}\Biggl[a^2\Delta_{VF}^{con}
(D_{xx}^V+D_{yy}^V)+2\biggl(-12m_{xy}^2-3q^2\sum_{a\in V}(1+\theta^{at})+
a^2\Delta_{VF}^{disc}\biggr)D_{xy}^V\nonumber\\
&+(V\rightarrow A)+a^2\Delta_{IF}^{con}(D_{xx}^I+D_{yy}^I)+2(3m_{xy}^2+a^2\Delta_{IF}^{con})D_{xy}^I\Biggr]
\label{disc444}
\end{align}
where
\begin{equation}
\Delta_{BF}^{disc}\equiv\sum_{a\in B}\biggl(\Delta^\prime_{at}+\theta^{at}\Delta_t - (3+2\theta^{at})\Delta_a\biggr)
\label{DeltaDBFDef}
\end{equation}
and $t\in F$.
Eq.~(\ref{disc444}) reduces to the result of Ref.~\cite{Aubin:2003mg}.
\begin{table}[h]
\begin{center}
\begin{tabular}{c || c | c | c | c | c }
\hline
B \slash F & V & A & T & P & I\\
\hline
V & $3C_1+6C_4+3C_6$ & $C_1+6C_4+9C_6$ & $2C_1+8C_4+6C_6$ & $0$ & $4C_1+12C_6$\\
A & $C_1+6C_3+9C_6$ & $3C_1+6C_3+3C_6$ & $2C_1+8C_3+6C_6$ & $0$ & $4C_1+12C_6$\\
I & $-C_3-3C_4$ & $-3C_3-C_4$ & $-2C_3-2C_4$ & $0$ & $-4C_3-4C_4$\\
\hline
\end{tabular}
\end{center}
\caption{The $\Delta_{BF}^{disc}$ result. multiply the coefficient $-6\times16/f^2$.}
\label{DiscDiag}
\end{table}

Taking into account the hairpin couplings, taste splittings, and
coefficients $\delta_{BF}^{con}$, $\Delta_{BF}^{con}$, and
$\Delta_{BF}^{disc}$ in Tables~\ref{ConDiag2}, \ref{ConDiag}, and
\ref{DiscDiag}.
We see that the results in Eqs.~(\ref{fin_con444}) and (\ref{disc444})
are invariant under the symmetry $V\leftrightarrow A$.

Collecting the analytic contributions to the self-energies from
Gasser-Leutwyler analytic terms and (\ref{eq:SV}) gives,
\begin{align}
\sigma^{anal}= &\phantom{+}\frac{16}{f^2}(2L_6-L_4)m_{xy}^2\,
4(m_{U_5}^2+m_{D_5}^2+m_{S_5}^2)+\frac{16}{f^2}(2L_8-L_5)\,m_{xy}^4\nonumber\\
&+\frac{16}{f^2}a^2(E^tm_{xy}^2 + F^t\,4(m_{U_5}^2+m_{D_5}^2+m_{S_5}^2)+ G^ta^2)
\label{eq:sanal444}
\end{align}
where we have absorbed terms proportional to $a^2\Delta_t$ into the
coefficients $E^t$, $F^t$, and $G^t$.
The first two terms of Eq.~(\ref{eq:sanal444}) correspond to the
continuum result and are the same for all tastes.
In the last three terms, the coefficients $F^t$ are degenerate within
irreps of $SO(4)_T$, while the coefficients $E^t$ and $G^t$ are
degenerate within irreps of $SW_\mathrm{3}$.
The exact chiral symmetry implies that $F^5=G^5=0$.  
Setting $t=\xi_5$ in Eq.~(\ref{eq:sanal444}) then yields the result of
Ref.~\cite{Aubin:2003mg}.

\subsection{\label{sec:res111}Results in 1+1+1 theory}
The results in Eqs.~(\ref{fin_con444}), (\ref{disc444}), and
(\ref{eq:sanal444}) must be modified to account for (partial)
quenching~\cite{Bernard:1992mk,Bernard:1993sv} and the fourth root of
the staggered fermion determinant \cite{milc-rmp-1}.
The replica method of Ref.~\cite{Damgaard:2000gh} allows us to
generalize to the partially quenched case.

The effects of partial quenching and rooting in
Eqs.~(\ref{fin_con444}), (\ref{disc444}), and (\ref{eq:sanal444}) are
easily summarized:
The quark masses $m_x$ and $m_y$ are no longer degenerate with members
of the set $\{m_u,m_d,m_s\}$, a factor of $1/4$ is introduced in the
second term of Eq.~(\ref{fin_con444}).
In addition, the $\delta_a$ of the mass matrices in
Eq.~(\ref{444_mass_matrix}) are replaced with the
$\frac{\delta_a}{4}$.
And terms in Eq.~(\ref{eq:sanal444}) that are proportional to the sum
of the sea quark masses are multiplied by $1/4$.

\section{Conclusion}
Tree-level contributions from the NLO Sharpe-Van de Water Lagrangian
break taste to the lattice geometric time-slice
group~\cite{Golterman:1985dz}.
Referring to Eq.~(\ref{eq:sanal444}) and noting the degeneracies of
the coefficients, we see that there are 21 \textit{a priori}
independent couplings entering these NLO analytic terms.
The three LECs appear in the Goldstone sector and the remaining 18
coefficients show up in the non-Goldstone sector.

The punch line is that the NLO chiral log terms respect the SO(4)
symmetry and this symmetry is broken to $\text{SW}_3$ only by those
analytic terms of Sharpe and Van de Water.
Hence, we can determine their LECs by simple subtraction between two
different irreps of $\text{SW}_3$, which belong to the same irrep in
SO(4).

\section{Acknowledgments}
The research of W.~Lee is supported by the Creative Research
Initiatives Program (3348-20090015) of the NRF grant funded by the
Korean government (MEST).

\end{document}